\documentclass[12pt,singlespace,oneside]{article}
\usepackage{amsmath}
\usepackage{latexsym}

\begin{document}

\title{Opera's neutrinos and the Robertson test theory of the Lorentz
transformations\thanks{%
To Dr. Douglas G. Torr, supporter and collaborator of several decades.} \ }
\author{Jose G. Vargas\thanks{%
PST\ Associates. 138 Promontory Rd., Columbia, SC 29209. USA} \ }
\date{}
\maketitle

\begin{abstract}
The difference in light's travel time from CERN to GPS to Gran Sasso, on the
one hand, and light going the direct route in vacuum (mimicked by
neutrinos), on the other hand, is analyzed with a modified Robertson test
theory of the Lorentz transformations. The modification consists simply in
removing the restriction of what Robertson referred to as agreement to
equate the to and from speeds of light. For reasons that will be presented
in a paper to soon follow, we restrict ourselves, within the new freedom, to
the case of preferred frame kinematics with absolute simultaneity.

At the level of not assuming any concomitant dynamical changes in this
alternative, the analysis yields \textit{zero effect}, i.e. no change with
respect to special relativity (to be expected). The 60 ns would thus remain
unexplained. However, a gravitation related effect that would likely
accompany an alternative kinematics yields that value up to uncertainties
due to the need to simplify the experimental set up for analysis. The effect
amounts to $(\lambda /2)(D/c)(V/c)^{2}$, where $D$ is distance to GPS, $V$
is speed with respect to the frame of isotropy of the cosmic background
radiation and $\lambda $ is a factor greater than $1$, but likely not
greater than $1.2$ or $1.3$, that reflects lack of precise knowledge of the
average distance to the common view GPS satellite. A $\lambda $ factor of \ $%
1.2$ yields 60 ns.
\end{abstract}

\section{Introduction}

In their analysis \cite{Opera}, the Opera group assumed the constancy of the
one-way speed of light, assumption shared by the physics community at large.
But most physicists and philosophers interested in the foundations of SR
assert that synchronizations and the one-way speed of light are matters of
convention covered under the umbrella of Reichenbach's thesis of
conventionality of synchronizations \cite{Reichenbach}. Consistently with
this, they maintain that certain transformations alternative to the ones by
Lorentz, like the para-Lorentzian transformations that we shall later
consider, do not actually represent a new physics. High energy physicists,
to name only the most obvious case, would find such a statement difficult to
digest (This author was caught in the defense of his Ph. D. dissertation
between a conventionalist and a high energy physicist; yes, I published and
got my degree!). Those observations and\ my little anecdote are meant to ask
for patience by skeptic readers, since the issue is a very subtle one. See
in this respect Chapter 8 of Bohm's book on the special theory of relativity %
\cite{Bohm}

Central to our discussion will be the realization that the present version
of the principle of equivalence is by itself enough to imply the Lorentz
transformations (LTs) without any such conventions \cite{Cattaneo} and,
therefore, imply a constant one-way speed that is not conventional (See in %
\cite{Cattaneo} references to previous related works, starting as early as
in 1910). There is no way out of this under present interpretations of the
relativity principle.

If the one-way speed of light were a matter of convention, the implications
of the now famous $60$ $ns$ should not be a cause for concern. A convention
(a definition for Einstein; an agreement for H. P. Robertson) cannot have
foundation-shaking consequences. How would neutrinos know what convention
was used to determine when they should arrive in Gran Sasso? However, there
is some sense in which one could speak of convention, though it is dangerous
to do so. I shall explain by advancing the contents of section 5.

If light traveled with a direction dependent speed different from $c$ from
CERN to GPS satellites and from there to Gran Sasso, so would the speed of
light in a straight tunnel (vacuum having been made in it) between CERN and
Gran Sasso. Under present understanding that we do not reject, neutrinos
would also travel with a speed slightly smaller but virtually equal to the
speed of light in vacuum. That speed would thus be, like light's under the
thesis of this paper, direction dependent. So, this measurement amounts in
principle to sending light between two places through direct and indirect
routes. Any changes in the arrival time of light from CERN to Gran Sasso via
GPS would be compensated exactly by the change in the arrival time of light
going the direct route in a straight tunnel. On the basis of results like
this, some may find it justifiable to make the assumption of equality of the
to and fro speeds since there is no way to tell the difference if it were
not so.

We shall consider an alternative with preferred frame that we shall call
para-Lorentzian (PL), an alternative that, either nameless or under a
different name, has been invoked by different researchers for various
purposes. It would be premature to give here acknowledgements, mainly
because of the lengthy and negative comments that would also have to be made
in more than one case. Future Opera-like experiments will tell whether the
subject is worth revisiting and assign credits, or discredits for that
matter.

The key part of this paper is sections 5 to 7. In section 5, we perform a
para-Lorentzian analysis of the Opera experiment under the trite assumption
that one should look at the kinematics only. It does not explain why
neutrinos appear to show up earlier than expected.

In section 6. we imagine what gravitation could be like in a
non-relativistic world that looks relativistic as per the preceding section,
but where we let Newton's law be directly related to the preferred frame,
rather than to the rest frame of the mass creating the gravitational field.
In other words, we are speaking of a potential gravitational concomitant of
a physics on the PL scenario.

In section 7, we compute the time of arrival of neutrinos in the context of
that concomitant. It agrees so closely with the $60$ $ns$ that luck must be
at work given the simplification of the experiment made to be able to
perform the analysis with available data and resources.

The rationale behind the contents of sections 6 and 7 will not be easily
understood without a paper which should soon follow \cite{V47}. It was
written a year ago as an outgrowth of a talk given a year ago at Bauman
University in Moscow. It makes the case for propertime as a fifth dimension.
The argument is based on superseding the foundations of present day
differential geometry. A new perspective emerges on issues such as the
equivalence principle. In section 8, the last one, we shall give a preview
of the implications of the 5-D scenario presented in that preprint \cite{V47}%
. Just to provide some limited perspective for the present paper, let us
just say the following. The LTs might be the facade of a more complex
reality consistent with absolute simultaneity and containing a preferred
frame in its fold; physics will continue to look relativistic and one will
continue to study it even if the physics presented in this paper were
correct and accepted. But there may be the occasional exception under
uncommon circumstances, Opera possibly being a case in point.

We complete the description of the contents of the paper. In section 2, we
show contradiction between a statement by Einstein about the one-way speed
of light and what his first postulate for SR implies. This contradiction
could not have been noted at the time. In section 3, we revise the Robertson
test theory of the Lorentz transformations \cite{Robertson} by removing from
it the equality of the two and fro speeds of light. In section 4, we deal
with the PL transformations. All that is necessary, or at least very
relevant, for the sections that then follow, whose contents has already been
presented in this introduction.

\section{Einstein on the one-way speed of light}

The opinion of Einstein on the one-way speed of light is retrospectively
relevant if we assume that it is not a constant, but that the two-way speed
is. If, for whatever reason, we set the one-way speed to be equal to the
two-way speed, the constancy of the latter implies the constancy of the
former. In this regard, Einstein in 1905 incurred in a contradiction that
could not be ascertained at the time, and that he felt might be possible:
``... We assume that this definition of synchronism is free from
contradictions ...'' \cite{Einstein}

The contradiction is stated using this and the next two paragraphs. Einstein
postulated the equivalence of all the inertial frames in spacetime. It was
not realized at the time that this suffices to obtain the LTs up to the
value of a constant, if negative (\cite{Cattaneo}, and references therein).
This implies that the square root of the negative inverse of that constant
is an invariant with dimension of velocity. This invariant is the only speed
which remains constant under those transformations. The one-way speed of
light is not then a matter of conventions, but to be determined by
experiment. The only reasonable option for that constant is the speed of
light. The only way out of this consequence of invariant one-way speed would
consist in not assuming the equivalence of all the inertial frames in the
usual sense, the only sense presently available.

The issue of synchronizations and of the constant speed of light are
intertwined. In paragraphs preceding immediately the aforementioned
citation, Einstein had stated: ``We have not defined a common ``time'' for A
and B, for the latter cannot be defined at all unless we establish \textit{%
by definition} that the ``time'' required by light to travel from A to B
equals the ``time'' it requires to travel from B to A ...'' Markings and
emphasis are as in the original. We wish to call attention to the``\textit{%
by definition}''.

Einstein defines times $t_{A}$, $t_{B}$ and $t_{A}^{\prime }$ instrumental
in stating the equality of the two and fro speeds of light as follows: ``In
accordance with definition the two clocks synchronize if $%
t_{B}-t_{A}=t_{A}^{\prime }-t_{B}$''. A definition is a convention. We
cannot make Plank's constant be whatever we want it to be as a matter of
definition. But Einstein did that with the one-way speed of light, even
though his postulate of equivalence implied that it is not a matter of
definition. End of explanation of contradiction.

Up to this point in this paper, the one way of speed of light has emerged as
being a direct consequence of a \textit{convention }\cite{Reichenbach}, a 
\textit{theorem} \cite{Cattaneo}, and a \textit{definition} \cite{Einstein}.
The issue then becomes what does the available experimental evidence say
about all this. Robertson wrote a most significance paper about it titled
``Postulate \textit{versus }observation in the Special Theory of
Relativity'' (emphasis in original). In the process of understanding its
essence, we shall again find the characterization of the one-way speed of
light\textit{. }We need to understand the essence of that paper.

\section{Revision of Robertson's test theory of the Lorentz transformations}

Robertson postulated \cite{Robertson}

(a) the existence of ``a reference frame $\Sigma $ ---Einstein's ``rest
system''---in which light is propagated rectilinearly and isotropically in
free space with constant speed c'' (Emphasis in original).

He assumed for his purpose

(b) that we may ``... confine ourselves to the consideration of events E in
a space-time neighborhood of a given event E$_{0}$ which is so small that we
may linearize $T$,...'' \ $T$ is the general transformation that constitutes
his starting point.

He then went to

(c) ``... \textit{adopt the procedure} proposed by Einstein for setting
clocks which are carried by observers $R$ at various points $x^{a}$ in the
reference frame $S$ '' (emphasis added). Robertson had previously introduced 
$S$ as Einstein's moving system.

In connection with synchronizing clocks at a distance through reflection of
a light signal and dividing by two the time of the two way trip, Robertson
writes:

(d) ``We \textit{agree} to set the auxiliary clock ... in such a way that it
records the time $t_{0}$ for the event E of reflection''.

Further down the text, he goes on to say

(e) ``... Einstein's synchronization insures \textit{as a matter of
definition }the equality of the forward and backward velocity along any
given line in $S$, ...'' Emphasis in original.

He interprets the null result of the Michelson-Morley (MM) experiment as
meaning:

(f) ``\textit{The total time required by light to traverse in free space, a
distance l and to return is independent of direction}''. Emphasis in
original.

He proceeds similarly with the experiment of Kennedy-Thorndike (KT):

(g) ``\textit{The total time required by light to traverse a closed path in }%
$S$ \textit{is independent of the velocity }$v$ \textit{of }$S$ \textit{with
respect to }$\Sigma $''. Emphasis in original.

And finally, for the Ives-Stilwell experiment (IS):

(h) ``\textit{The frequency of a moving atomic source is altered by the
factor (}$\mathit{1}$\textit{-}$u^{2}/c^{2}$\textit{)}$^{1/2}$\textit{,
where }$u$ \textit{is the velocity of the source with respect to the
observer.'' }Emphasis in original.

In that way, Robertson reaches the standard metric of special relativity. As
is well known, it is left invariant by translations, LTs, spatial rotations
and products of those.

If one gives a value to the one way speed of light in the path from CERN (C)
to GPS (S) to Gran Sasso (G), one should ignore (c), (d) and (e) and treat
Opera as a fourth experiment ---though highly imperfect because of its many
complications--- for a the revised Robertson test theory of the LTs. To be
precise, the analysis will induce us into some back engineering that it
takes us beyond the testing of the LTs. We are led into whether we can
consider an alternative to the LTs in a purely kinematical context. Under a
plausible concomitant change in Newton's law at the level of one part in $%
10^{6}$ for the earth, Opera's non-null result might be a way in which
nature might be voicing lack of conformity with (c).

In discussing the transformations that we are about to consider in the next
section, Thirring \cite{Thirring} stated ``... it would be possible to save
the notion of absolute simultaneity, but no one is willing to make the
necessary \textit{sacrifices any more}: \ philosophical principles are not
as persuasive as \textit{mathematical elegance}" (emphasis added).

To the key terms \textit{convention, theorem} and \textit{definition }we add
the also key terms \textit{adoption of a procedure, agreement, sacrifices
any more, }and \textit{mathematical elegance.}

In the present paper, the one-way speed of light is a theorem if one assumes
the only version of the equivalence principle that appears to be possible in
a pure spacetime context. In the remainder of this paper, we shall present a
new context that will be further developed in the announced sequence \cite%
{V47}, more concerned with mathematical elegance.

\section{Para-Lorentzian transformations}

The revision of the Robertson test theory is better performed in terms of
coordinate transformations, provided they represent changes of frames and
not just simply changes of markers (See paragraph of Eq. (6)).

Of all the members of the family of transformations that comply with (f),
(g) and (h) but not (e), we are now choosing the specific member where time
is absolute. This is not only for esthetic or simplicity reasons, but
because PL\ is the only member of the family chosen by our theoretical
considerations on the arena of physics \cite{V47}. These transformations
have been considered by a few others in the past but not worth mentioning at
this point because negative comments should also accompany what may
otherwise be potentially important contributions. But, if these
transformations prove their worth, that work should revisited.

Let $\mathbf{V}$ be the velocity of the laboratory system with respect to $%
\Sigma (T,X,Y,Z).$ Its components can be chosen to be $(v,0,0)$ with respect
to some appropriately oriented axes for $S^{\prime }(t^{\prime },x^{\prime
},y^{\prime },z^{\prime })$ . With the Robertson approach but with the
specified change, one can easily show how to mathematically represent a
hypothetical non-compliance of nature with any one or several of (e), (f),
(g) and (h) \cite{V9}$\mathbf{.}$ The family of ``transformations'' that
replaces the LTs if (d) is not satisfied is%
\begin{equation}
x^{\prime }=\gamma \cdot (X-VT),\text{ \ \ }y^{\prime }=Y\text{, \ \ }%
z^{\prime }=Z\text{, \ \ }t^{\prime }=hX+(\gamma ^{-1}-hV)T,  \tag{1}
\end{equation}%
where $h$ is a family of functions of $V$. These transformations contain
both the LTs and the PL transformations, the last ones being%
\begin{equation}
x^{\prime }=\gamma \cdot (X-VT),\text{ \ \ }y^{\prime }=Y\text{, \ \ }%
z^{\prime }=Z\text{, \ \ }t^{\prime }=\gamma ^{-1}T.  \tag{2}
\end{equation}%
We define unprimed low case quantities by%
\begin{equation}
x=x^{\prime },\text{ \ }y=y^{\prime },\text{ \ }z=z^{\prime },\text{ \ \ }%
t=t^{\prime }-Vx^{\prime }  \tag{3}
\end{equation}%
and use (3) in (2). We thus identify $S(t,x,y,z)$ as a relativistic frame
with the same velocity as the PL frame $S^{\prime }$ since%
\begin{equation}
x=\gamma \cdot (X-VT),\text{ \ \ }y=Y\text{, \ \ }z=Z\text{, \ \ }t=\gamma
(T-Vx).  \tag{4}
\end{equation}%
For arbitrary direction of the velocity, the relation between $S$ and $%
S^{\prime },$%
\begin{equation}
\mathbf{r}=\mathbf{r}^{\prime },\text{ \ \ \ \ \ \ \ }t=t^{\prime }-(\mathbf{%
V\cdot r),}  \tag{5}
\end{equation}%
allows us to compute more easily than resorting directly to $\Sigma .$

All these coordinate transformations are to be viewed not as changes of
markers but as representative of changes of frames. In affine space, the
equation%
\begin{equation}
x^{\mu }\mathbf{e}_{\mu }=x^{\prime \mu }\mathbf{e}_{\mu }^{\prime }, 
\tag{6}
\end{equation}%
expresses that the translation vector can be expressed in terms of just any
vector basis. A change of coordinates thus induces a change of basis and
vice versa. But alternative coordinates can also be viewed simply as
different sets of markers. This comment also applies to changes between
Lorentz frames and frames for kinematics that do not comply with any one or
several of (d), (e), (g) and (e). Hence, there is nothing special about (5)
in this specific regards; one could do something totally similar for any
kinematics that does not satisfy any of Robertson's three optical
experiments.

The preceding considerations have an important consequence. One is not going
to find any effects in experiments where there is not some element that
distinguishes a change of basis from a change of markers, as the
computations would be the same in both cases; since no effect can arise from
a change of markers, no effect would be found by identical computations for
the change of frames, if there is not such a distinguishing element.

Returning to our computations in a PL world, let $\mathbf{c}$ and $\mathbf{c}%
^{\prime }$ be the velocity of light relative to $S$ and $S^{\prime }.$ We
have%
\begin{equation}
\mathbf{c}^{\prime }=\frac{d\mathbf{r}^{\prime }}{dt^{\prime }}=\frac{d%
\mathbf{r}^{\prime }}{dt}\frac{dt}{dt^{\prime }}=\frac{d\mathbf{r}}{dt}\frac{%
dt}{dt^{\prime }}=\mathbf{c(}1-\mathbf{V\cdot c}^{\prime }),  \tag{7}
\end{equation}%
where we have used (4). Since $\mathbf{c}$ and $\mathbf{c}^{\prime }$ are
colinear, their unit vectors are equal 
\begin{equation}
\mathbf{c=n}=\mathbf{n}^{\prime }.  \tag{8}
\end{equation}%
where $c^{\prime }\equiv \left| \mathbf{c}^{\prime }\right| $ and where we
take $c$ as unit of speed. For $V\ll 1$, we get from (7),%
\begin{equation}
\frac{1}{c^{\prime }}=1+\mathbf{V\cdot c}^{\prime }+(\mathbf{V\cdot c}%
^{\prime })^{2}+...  \tag{9}
\end{equation}%
We also have\textbf{\ }%
\begin{equation}
\mathbf{c}=\frac{d\mathbf{r}}{dt^{\prime }}\frac{dt^{\prime }}{dt}=\mathbf{c}%
^{\prime }\mathbf{(}1+\mathbf{V\cdot c}),  \tag{10}
\end{equation}%
and%
\begin{equation}
\frac{1}{c^{\prime }}=1+\mathbf{V\cdot c=}1+\mathbf{V\cdot n=}1\mathbf{%
+V\cdot n}^{\prime },  \tag{11}
\end{equation}%
which are exact equations.

\section{Para-Lorentzian analysis of the Opera measurement}

We shall now proceed to present an analysis of the problem of measurement of
the one way speed of light tailored-made to the Opera experiment. This is a
refinement and generalization of a similar argument Bohm made in just one
dimension \cite{Bohm}.

In $S^{\prime }$, the position vectors from C to S, from S to G and from C
to G will be denoted as $\mathbf{D}_{C}$, $\mathbf{D}_{G}$ and $\mathbf{d}$.
The respective unit vectors will be named $\mathbf{n}_{_{C}}^{\prime }$, $%
\mathbf{n}_{_{G}}^{\prime }$and $\mathbf{n}^{\prime }\mathbf{_{_{\nu }}.}$
And we shall use the symbols $c_{_{C}}^{\prime }$, $c_{_{G}}^{\prime }$ and $%
c^{\prime }\mathbf{_{_{\nu }}}$ to refer to the speed of light corresponding
to those directions. Obviously%
\begin{equation}
\mathbf{D}_{C}+\mathbf{D}_{G}-\mathbf{d=0.}  \tag{12}
\end{equation}%
We shall now show that light satisfying equations (7) to (11) take the same
time to cover the closed path CSGC (using a hypothetical tunnel between C
and G) as if light were travelling always at constant speed $c.$ This time
is given by%
\begin{equation}
\frac{D_{C}}{c_{_{C}}^{\prime }}+\frac{D_{G}}{c_{_{G}}^{\prime }}+\frac{d}{%
c^{\prime }\mathbf{_{_{-\nu }}}}\text{ }\mathbf{=}\text{ }0\mathbf{,} 
\tag{13}
\end{equation}%
where $c^{\prime }\mathbf{_{_{-\nu }}}$ represents the speed of light if it
were going from G to C.

Using (11), expression (13) becomes%
\begin{equation}
\frac{D_{C}}{c}(1\mathbf{+V\cdot n}_{_{C}}^{\prime })+\frac{D_{G}}{c}(1%
\mathbf{+V\cdot n}_{_{G}}^{\prime })+\frac{d}{c}(1-\mathbf{V\cdot n}_{_{\nu
}}^{\prime })\text{ }\mathbf{=}\text{ }0\mathbf{,}  \tag{14}
\end{equation}%
where we have made $c$ explicit outside the parenthesis to make obvious that
we are adding time intervals. The sum of the terms obtained from the units
in the parentheses yields the time that light would take to cover the path
with speed $c$. The other terms yield the dot product by $\mathbf{V}$ of the
left hand side of (12), and is thus null. The point of null effect has been
made.

Notice that the result just obtained is exact and, therefore, independent of
how large $c^{\prime }$ could become. And it is because cancellations like
this are the norm rather than the exception that one should not be concerned
about an underlying reality that were not relativistic; it might still
appear to be so.

We proceed to develop another important observation in order to understand
where we should look or not look for 60 ns in the present context. Let $t$
and $t^{\prime }$ be the times taken by light to cover the trajectories
indicated by subscripts. We have proved that 
\begin{equation}
t_{_{CSG}}^{\prime }+t_{_{GSC}}^{\prime }=t_{_{CSG}}+t_{_{GSC}},  \tag{15}
\end{equation}%
and similarly 
\begin{equation}
t_{_{CG}}^{\prime }+t_{_{GC}}^{\prime }=t_{_{CG}}+t_{_{GC}}.  \tag{16}
\end{equation}%
Solving for $t_{_{GC}}^{\prime }$ in (16) and substituting in (15), we get%
\begin{equation}
t_{_{CG}}-t_{_{CG}}^{\prime }=t_{_{CSG}}-t_{_{CSG}}^{\prime }.  \tag{17}
\end{equation}%
$t_{_{CG}}$ is the time that neutrinos take to go from C to G at the speed $%
c $, and $t_{_{GC}}^{\prime }$ is the actual time they take as per PL.

Notice that the left hand side of equation (17) does not depend on the point
S. Hence neither can the right hand side even if each of its two terms does.

To conclude, if PL\ had to do only with the dependence of the speed of light
with direction, it could not account for the effect. Many would be saying: I
told you so. But there is far more to spacetime related structure in PL than
just the speed of light. See, for instance, Maciel and Tiomno in connection
with the concept of (practical) rigidity in SR and PL \cite{MacTio}. And 
\textit{if there were more}, a law of nature would determine when neutrinos
should arrive in Gran Sasso, not some convention that neutrinos do not know
of.

\section{Newton's law in preferred frame context}

As explained by Maciel and Tiomno, the rotation of a practically rigid body
defines two different physical configurations depending of whether we are in
a SR world or a PL world. The role that practical rigidity plays in the
analysis by Maciel and Tiomno of resonance experiments in centrifuges \cite%
{MacTio} is played by Newton's law in the Opera experiment. In one case as
in the other, it is a matter of the differences ab initio between physical
configurations of an experiment which, against all appearances, is not the
same in a world as in the other, say the worlds of SR and PL.

Equations (5) by themselves would seem to preclude any such difference,
since Newton's law pertains to gravitostatics, where only displacement
vectors in three dimensions should matter; these are the same in SR as in
PL. Nature, however, could be more subtle. We have to imagine what such a
PL\ world might look like.

Here is the crux of the problem. Given the peculiarities of light (as per
the previous section), synchronizations that take place in actual life are
Einstein's, not PL synchronizations. Consider now slow clock transport.
Unless a clock is at rest in $\Sigma $, it is not moving slowly. Only if the
laboratory were almost at rest in $\Sigma $ would slow clock transport be a
valid, approximate synchronization procedure in a $PL$ world, but the reason
would be that, in that frame, SR and PL physics would be practically
identical. In a PL world there is an \textit{unavoidable disconnect }between
its \textit{hypothetical relation of simultaneity }and how clocks actually
behave. Because of this, SR will always be part of the physics, barring the
exceptional situations that catch this ``malicious nature'' off guard.

Synchronizations are key to the functioning of the GPS system but, in the
weak or virtually flat approximation of the preferred frame scenario,
Newton's dynamical law would take its standard form in the frame $\Sigma $,
not in the frame $S^{\prime }.$ That is the thesis we use here to analyze
Opera. Given that Newton's constant $G$ and the mass of the earth are known
only to one part in $10^{4}$, the highly precise \textit{GPS system is an
almost miraculous feat} of scientific and technological engineering. It is
monitored and corrected continuously, using the more reliable system of
atomic clocks on earth. Thanks to the \textit{stability provided through
that monitoring }and the self-consistency between the participant satellites
that one so achieves, one can measure small distances and achieve meaningful
synchronizations using the two-way speed of light.

One computes highly precise differences between the position vectors of two
places without subtracting the would-be-imprecise actual distances to the
center of the earth. In addition, one is dealing with distances computed
with intervention of two-way speeds of light. This is not obvious because,
although one-way ranging is involved in the common use of the GPS system,
the basic functioning of the system involves satellites viewed in different
directions. It has the same character as two-way determinations. All this
goes to say that these \textit{computed measurements amount to corrections
rather than actual values. }This is like the quantum field corrections of
the hydrogen atom's fine structure. Relativistic quantum mechanics provides
the bulk of the value (so to speak), but not the corrections; quantum field
theory on its own cannot provide the value, only the corrections. Hence it
does not matter for most purposes whether we are working with a very precise
value for the mass of the earth or not.

From a PL perspective, the GPS system would be a hybrid one. The foundations
of its gravitational dynamics would not be in accordance with the
synchronizations that actually take place. The foundations are ``preferred
frame based'', \textit{but measurements with this system (say distances in
the Opera experiment) are differential \ }and thus based on local
assumptions related to the wrong synchronizations. For each neutrino event,
the distance covered by light in going from CERN to GPS satellite to Gran
Sasso is determined using just one satellite, in common view from both
places. It is an apparent $S^{\prime }$ distance because it is computed
through the use of signals from a self-consistent system but whose orbit is
then computed by ```wrongly using'' the underlying Newton's law. It should
be based on $\Sigma $ distances, not on $S^{\prime }$ distances, much less
on apparent $S^{\prime }$ distances.

Consequently, we proceed to compute the effect on the time of flight of
light under a presumed effect of this unavoidable mismatch. We assume that
the real distance in $S^{\prime }$ is not the one actually considered to
have been measured in $S^{\prime }$ (through computation on the basis of the
wrong assumptions), but that it is related to it like the real $\Sigma $
distance is related to the the real $S^{\prime }$ distance. Thus we shall
use the relation between $\Sigma $ and $S^{\prime }$ for our analysis. How
could one do better than this? There are three legs to the answer:
theoretical, phenomenological and practical.

From a theoretical perspective, one cannot do better without developing the
beginning of a theory that covers at least a new gravitation the facade of
which (loosely speaking) is Einstein's theory of gravity.

From a practical perspective, a PL-consistent GPS\ system would be
practically unworkable. Those who designed the system already know what is
practical because, as explained, they developed a very precise system going
beyond the precision in the knowledge of $G$ and of the mass of the earth.

From a phenomenological/modelling perspective, a more sophisticated analysis
would have to be carried out by those who designed the system and maintain
it, but perhaps only for a case like this, where one sees the frame behind
the facade. Such an effort is unwarranted at this point. Confirmation of the
same anomaly by other experiments should take place first, specially since
such additional experiments appear to be in the workings already.

Finally a word about something else, just in case. Inconsistencies have
appeared in measurements of $G$ at a level as high as 1 part in $10^{4}.$
The second order effect that we are about to consider would be too small in
order to explain such inconsistencies. But one might not be able to totally
exclude first order effects in other situations.

\section{Analysis of the Opera experiment with concomitant gravitational
effect}

Let us assume that we follow the course of action just outlined. We rewrite
Eqs. (2) as%
\begin{equation}
X=\gamma \cdot (x^{\prime }+Vt^{\prime }),\text{ \ \ }Y=y^{\prime }\text{, \
\ }Z^{\prime }=z^{\prime }\text{, \ \ }T=t^{\prime }.  \tag{18}
\end{equation}%
For arbitrary direction of velocity of $S^{\prime }$, the assignment of
3-vectors to pairs of ``points'' (meaning C and S, or S and G, or C and G)
at constant time is given by%
\begin{equation}
\Delta \mathbf{R}=\Delta \mathbf{r}^{\prime }+\mathbf{V}\frac{\gamma -1}{%
V^{2}}(\mathbf{V\cdot }\Delta \mathbf{r}^{\prime }),  \tag{19}
\end{equation}%
directly from the spatial part of the PL transformations, the same as for
the LTs. In second order%
\begin{equation}
\Delta \mathbf{R}=\Delta \mathbf{r}^{\prime }+\frac{1}{2}\mathbf{V}(\mathbf{%
V\cdot }\Delta \mathbf{r}^{\prime }).  \tag{20}
\end{equation}%
With $\Delta \mathbf{r}^{\prime }=\mathbf{D}=D\mathbf{n}^{\prime }$, we have%
\begin{equation}
\Delta \mathbf{R}=D\left[ \mathbf{n}^{\prime }+\frac{1}{2}V^{2}(\mathbf{n}%
^{\prime }\mathbf{\cdot n}_{_{V}})\mathbf{n}_{_{V}}\right] ,  \tag{21}
\end{equation}%
and, therefore,%
\begin{equation}
\left| \Delta \mathbf{R}\right| =D\left[ 1^{\prime }+V^{2}(\mathbf{n}%
^{\prime }\mathbf{\cdot n}_{_{V}})^{2}+...\right] .  \tag{22}
\end{equation}

Let us return to the specifics of the Opera measurement. $d$ is about 732
kms. $D_{C}$ and $D_{G}$ will be larger than $2\times 10^{4}$ kms for the
satellite in ``common-view from C and G'', but of that order. At best, only
for some very special configuration would the relative smallness of $d$ be
compensated by negligible trigonometric factors in the $D$ terms. Because of
the averaging over neutrino event(s), such special configurations lack
relevance, if there were any in the first place. In fact there is none given
the latitudes of M and of $\mathbf{V}$ (see next paragraphs). Hence, we may
neglect the $d$ term, set $D_{C}=D_{G}$ $\equiv D_{M}$, where M is the mid
point between C and G, and choose this value to be $2\times 10^{4}$ kms.

The orientation that the line of view of the satellite makes with the
vertical of the mid point M will be within a solid cone\ where the
generating line makes, say, 45$^{0}$ with the axis, the vertical at M. For
different events, the line of view will bell over the place within that
solid angle. So, a factor of 1.2 multiplying $2\times 10^{4}$ kms may be
justified, while at the same time replacing those satellites for
simplification purposes with a stationary hypothetical ``GPS station'',
which would certainly not be a satellite. Because of the relative smallness
of $d$, we can take the unit vectors $\mathbf{n}_{_{C}}^{\prime }$ and $%
\mathbf{n}_{_{G}}^{\prime }$ as $\mathbf{n}_{_{M}}^{\prime }$ and -$\mathbf{n%
}_{_{M}}^{\prime }.$ We thus have for the excess of time delay by light
going the GPS route%
\begin{equation}
2\frac{DV^{2}}{c^{3}}(\mathbf{\mathbf{n}_{_{M}}^{\prime }\cdot n}_{V})^{2}. 
\tag{23}
\end{equation}

The candidate $\mathbf{V}$ is the one in which the 2.7 degree cosmic
microwave background (CBM) radiation would be viewed as isotropic. Its
magnitude varies because of the translation of the earth around the sun. We
take the value given by Smoot \cite{Smoot} for the speed of the sun with
respect to CBM rest frame as approximate value of the direction of the speed
of the earth. That value is $368\pm 2$ kms per second. For the same reason
of neutrino events averaging mentioned above, the right ascension of $%
\mathbf{V}$ will be irrelevant, and we can take it to be zero, i.e. as
origin of redefined right ascensions. Its declination is -$\mathbf{7}^{\circ
}.$ Since the latitude of M is approximately $\mathbf{45^{\circ }}$, its
vertical will never be anywhere near the direction of $\mathbf{V}$.

The unit vector in the direction of $\mathbf{V}$ will then be taken to be%
\begin{equation}
\mathbf{\mathbf{n}_{_{V}}^{\prime }=(0,}\text{ }\mathbf{-\cos 7}^{\circ }%
\mathbf{,}\text{ }\mathbf{-\sin 7}^{\circ }),  \tag{24}
\end{equation}%
and, as the earth rotates,%
\begin{equation}
\mathbf{\mathbf{n}_{_{M}}^{\prime }=(\cos 45^{\circ }\cos \phi ,}\text{ }%
\mathbf{\cos 45^{\circ }\sin \phi }\text{, sin }\mathbf{45^{\circ }}). 
\tag{25}
\end{equation}%
The dominant term in $(\mathbf{\mathbf{n}_{_{V}}^{\prime }\cdot \mathbf{n}%
_{_{M}}^{\prime }})^{2}$ is ($\mathbf{\cos 7}^{\circ }\mathbf{\cos 45^{\circ
}\sin \phi )}^{2}$, whose average is approximately $1/4.$

We are thus left with $(1/2)(D/c)(V/c)^{2}$ excess time of flight of light.
This expression yields $50$ $ns$ for the given values of $D$ and $V$ . If
we\ replace $D$ with $1.2$ $D$ because the vertical is a reasonable average
for directions, but not for distances, we get $60$ $ns$. Of course, we could
have chosen $1.1$ or $1.3$ instead of $1.2$. The delay would then be $55ns$
or $65ns$. The coincidence is extraordinary; the theoretical argument, not
too solid yet.

It would readily occur to readers the question of what about the Pioneer
probes anomalies. Though one can never be sure of having done the correct
analysis in virgin territory, our analysis shows that, if there is an
effect, it is much smaller than the anomaly.

\section{Concluding remarks}

With regards to the rationale for five dimensions, consider the following.
The theory of connections is a theory of moving frames and only moving
frames, whether it is presented that way or not. For an easy to understand
example, let us mention that particles are left out of the core of Cartan's
theory of connections, which is the paradigm of local differential geometry:
they do not enter the equations of structure. If propertime ($\tau $) is
viewed as a fifth dimension, the dual of its differential, the
4-vector-velocity, is represented as having a projection on (but not as been
contained in) the spacetime subspace. Though not the most interesting
implication of five dimensions, particles will thus enter the equations of
structure through the\ ``4-velocity'' viewed as tangent vector dual to
proper time.

It will be claimed in the next paper, that $U(1)\times SU(2)$ is to the $%
x\tau $ subspace what the Lorentz group is to $tx$. Here, by Lorentz group,
we mean the actual group of the Lorentz transformations, or to the group of
the para-Lorentzian transformations (the general case does not look as the
very special case considered), which is a non-linear representation of the
Lorentz group \cite{Rembielinski}, \cite{V10}. But, if $U(1)\times SU(2)$ is
directly related to the tangent bundle, should $SU(3)$ not also be so?
Schmeikal \cite{Schmeikal1}-\cite{Schmeikal2} has shown that $SU(3)$ also
can be related to the tangent bundle, fermion density functions being
represented by primitive idempotents in the Clifford algebra of spacetime.
In the present author's opinion, the primitive idempotents should not be
those of the tangent bundle but of the K\"{a}hler algebra of differential
forms. A publication on this subject will see the light when the motivation
will be there.

Consider next the equivalence principle. The 4-velocity $\mathbf{u}$ is now
a vector spanning the fifth dimension but not orthogonal to spacetime. Its
projection on it has naturally components $\gamma (1,u^{i})$. In PL, on the
other hand, its projection is $\gamma ,0,0,0.$ In other words it is
orthogonal to the 3-space subspace as in SR. We still have an equivalence
principle which favors PL over SC. In our analysis \cite{V47}, the $x\tau $
subspace is relativistic if$\ $spacetime ($tx$) is PL, and it is
non-relativistic if $tx$ is $SR$-like.

Finally, consider the role we assigned to distances in $\Sigma $ for the
analysis of the Opera result. It will emerge that gravitation has to do with
universal time and universal distances. The other interactions will be
concerned with propertime and their 3-space rest frame. Electrodynamics has
double life, represented by $U(1)$ when (electromagnetic spinorial)
structures are concerned, and invariant under the Lorentz group when
particles in external fields are concerned.

To conclude, the thesis presented in the combination of this and the
following paper ---thesis to which Opera lends a hand--- is that the world
looks relativistic after all. However, given the exceptions, even if
presently very few, it is more accurate to say that SR is the\textbf{\ facade%
} of a more complex theory which can only be seen through cracks in the
facade. If the thesis advanced in this\ paper happens to be correct, SR and
relativistic quantum mechanics will continue to be taught ---and not simply
as an approximation like Newtonian mechanics is--- but in the context of a
very subtle, more complex reality.

\section{Acknowledgements}

Relevant conversations with Professors Zbigniew Oziewicz and Timir Datta,
and Doctors Bernd Schmeikal and Stanley Alterman are acknowledged. Help with
software from Dr. Iulian C. Bandac is appreciated.

\end{document}